\documentclass[aps,prl,twocolumn,showpacs,floatfix,superscriptaddress]{revtex4}

\usepackage{bm,bbm,amsmath, amssymb,amsbsy, amsthm,graphicx,subfigure,color,txfonts,multirow,mathtools}
\usepackage[framemethod=tikz]{mdframed}
\usepackage[colorlinks]{hyperref}
\hypersetup{
	bookmarksnumbered,
	pdfstartview={FitH},
	citecolor={blue}, 
	linkcolor={red},
	urlcolor={black},
	pdfpagemode={UseOutlines}
}

\usepackage{bm}
\usepackage{empheq}
 
\usepackage{upgreek}
\usepackage{cancel}
\usepackage{slashed}

\newcommand{\ket}[1]{\left| #1 \right\rangle}
\newcommand{\bra}[1]{\left\langle #1 \right|}

\newcommand{\Ignore}[1]{}

\renewcommand{\eqref}[1]{Eq.~(\ref{#1})}

 \newcommand{\tr}[1]{\text{Tr}}

\begin{document}

\title{Quantifying genuine multipartite correlations and their pattern complexity}

\author{Davide Girolami}
\email{davegirolami@gmail.com}
\affiliation{$\hbox{Department of Atomic and Laser Physics, University of Oxford, Parks Road, Oxford OX1 3PU, UK}$}
\author{Tommaso Tufarelli}
\affiliation{$\hbox{School of Mathematical Sciences, The University of Nottingham, University Park, Nottingham NG7 2RD, UK}$}
\author{Cristian E. Susa}
\email{cristiansusa@correo.unicordoba.edu.co}
\affiliation{$\hbox{Departamento de F\'isica y Electr\'onica, Universidad de C\'ordoba, Carrera 6 No. 76-103, Monter\'ia, Colombia}$} 
\begin{abstract}

We propose an information-theoretic framework to quantify multipartite correlations in classical and quantum systems,  answering  questions such as:  what is the amount
of  seven-partite correlations in a given state of ten particles? We identify  measures of  genuine multipartite correlations, i.e. statistical dependencies which cannot be  ascribed to bipartite correlations, satisfying a set of desirable properties. 
Inspired by ideas developed in complexity science, we then introduce the concept of weaving to classify states which display different correlation patterns, but cannot be distinguished by correlation measures. The weaving of a state is defined as the weighted sum of correlations of every order. Weaving measures are good descriptors of the complexity of correlation structures in multipartite systems.

  \end{abstract}

\date{\today}
       
 
  \maketitle

{\it Introduction --}
Statistical relations in measurement  outcomes, i.e. correlations, are powerful tools to investigate multipartite systems, employed in (quantum) information theory, statistical mechanics,  condensed matter theory, network theory, neuroscience, and complexity science \cite{badii,condmat,cover,wilde}. Correlations describe global properties which cannot be inferred from the features of the system parts, e.g. phases of many-body systems \cite{many}. They are also resources. Entanglement, a kind of quantum correlation, enables speed-up in quantum information processing \cite{ent}.   \\
Yet, the very notion of genuine multipartite correlations still generates discussion \cite{bennett}. There is no consistent way to quantify dependencies which do not manifest bipartite correlations, encoding joint properties of $k>2$ particles instead, while witnesses of multipartite entanglement of {\it at least order $k$} have been proposed \cite{multi1,multi2,multi3,multi4,multi5,multi6,multi7,szalay}. A further problem is that  computing correlations is not always sufficient to fully describe multipartite  correlation patterns. Equally correlated  networks of multivariate  variables can display different structures and properties \cite{comp1,comp2}.  Also, quantum states can be correlated in inherently inequivalent ways   \cite{cirac,acin,nine}. \\  
Here we propose a framework to describe genuine multipartite correlations in classical and quantum systems.  We identify distance-based measures which satisfy a set of desirable properties when parts of the systems are added or discarded, and local operations are performed. We show that adopting the relative entropy allows for simplifying computations and meeting even stronger constraints.  We then introduce the notion of  {\it weaving}  to classify multipartite states by studying how correlations scale with their order. The weaving of a state is given by the weighted sum of genuine multipartite correlations of any order, inheriting the properties of correlation measures.  We compute the weaving of correlated classical and quantum states. In such cases,  states which have equal total correlations or highest order correlations, but display a different correlation pattern, take different weaving values.  \\ 

\begin{figure}[t]
 \includegraphics[width=7cm,height=5cm]{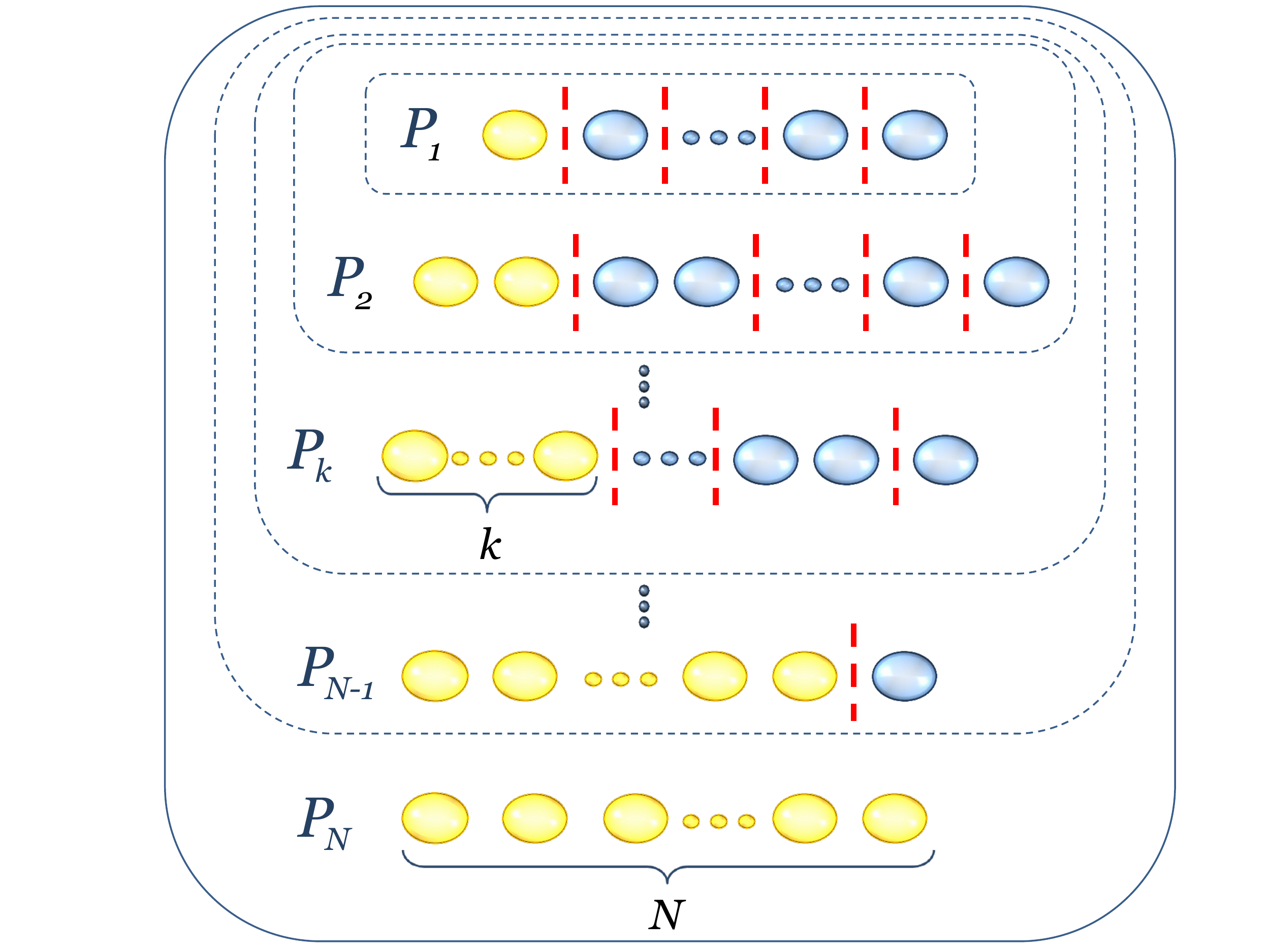}
 \caption{Multipartite correlation hierarchy. Given a system of $N$ particles (blue spheres),  the sets $P_k, k=1,2,\ldots,N,$ consist of states displaying up to $k$-partite correlations. The yellow $k$ spheres identify the largest subset of a coarse grained partition (the dashed red lines separate each cluster). The amount of genuine $k$-partite correlations in a state is the difference  between the distance to the sets $P_{k-1}$ and $P_k$.}
 \label{fig1} 
 \end{figure}

 {\it Quantifying Genuine Multipartite Correlations --}
A finite dimensional $N$-partite quantum system  ${\cal S}_{N}=\{{\cal S}_{[1]},{\cal S}_{[2]},\ldots,{\cal S}_{[N]}\},$ is described by a density matrix $\rho_{N}$, being $\rho_{[i]}$  the states of the subsystems ${\cal S}_{[i]}, i=1,2,\ldots,N$. In this framework,   classical probability distributions $p_{\alpha_1,\ldots,\alpha_N}$ of  $N$-variate discrete variables are embedded in density matrices of the form $\sum_{\alpha_1,\ldots,\alpha_N}p_{\alpha_1,\ldots,\alpha_N}\ket{\alpha_1,\ldots,\alpha_N}\bra{\alpha_1,\ldots,\alpha_N}, \sum_{\alpha_1,\ldots,\alpha_N}p_{\alpha_1,\ldots,\alpha_N}=1$, where  $\{\alpha_i\}$ are orthonormal basis elements in the Hilbert spaces of each subsystem ${\cal S}_{[i]}$. 
The correlations in ${\cal S}_N$ depend on the tensor product structure of its Hilbert space, induced by the  partition $\{{\cal S}_{[i]}\}$. This is usually dictated by  physical constraints, e.g. spatial separation of the subsystems. Indeed, even maximally entangled states are factorizable by changing the system structure \cite{viola,lloyd1}. 
The total correlations  in the  system  represent  the information encoded in $\rho_{N}$ which is unaccessible to an observer knowing the states of each subsystem, $\rho_{[i]}$. We extend the argument to define genuine multipartite correlations of order higher than $k, 2\leq k\leq N-1,$  as the missing information to a more informed observer, who knows the states  $\rho_{k_j}$ of clusters forming a coarse grained partition $\{{\cal S}_{k_1}, {\cal S}_{k_2},\ldots, {\cal S}_{k_m}\},  \sum_{j=1}^m k_j=N, k_j\leq k$, where each cluster ${\cal S}_{k_j}$   includes up to  $k$  subsystems,  e.g. ${\cal S}_{k_1}=\{{\cal S}_{[1]},{\cal S}_{[2]},\ldots, {\cal S}_{[k_1]} \}$. Genuine $N$-partite correlations, the highest order,  are the information that is still missing when clusters including subsets of up to $N-1$ subsystems are accessible, $k_j\leq N-1$. The set of states describing clusters of up to $k$ subsystems  is
\begin{eqnarray}\label{pk}
P_{k}:=\left\{ \sigma_N=\otimes_{j=1}^m \sigma_{k_j}, \sum_{j=1}^m k_j=N, k=\max\{k_j\} \right\}.
\end{eqnarray}
For example, given $N=3$, the set $P_{1}$ consists of  the product states $\otimes_{j=1}^3\sigma_{[j]}$, $P_{2}$ includes $P_{1}$  and  the products of bipartite and single-site states, i.e. $\sigma_{2}\otimes\sigma_{1}$ and their permutations, while $P_{3}$ contains $P_{2}$ and the non-factorizable density matrices $\sigma_{3}$. The  complete chain reads $P_{1}\subset P_{2}\subset\ldots\subset P_{N-1}\subset P_{N}$, where $P_{N}$ is the Hilbert space of  the global system   (Fig.~\ref{fig1}). Note that a pure state in $P_k$ is a $k$-producible state \cite{briegel}.
Genuine multipartite correlations of order higher than $k$  are then quantified  by the distance of the global state to the set $P_k$,
\begin{equation}\label{min}
D^{k\rightarrow N}(\rho_N):=\min\limits_{\sigma\in P_{k}}D(\rho_N,\sigma),
\end{equation}
where the function $D$ is  non-negative, $D(\rho,\sigma)\geq0, D(\rho,\sigma)=0\iff \rho=\sigma,$ and contractive  under completely positive and trace preserving (CPTP) maps $\Phi,     D(\rho,\sigma) \geq D(\Phi(\rho),\Phi(\sigma)), \forall \rho,\sigma$. 
Then, any distance  identifies a measure of  $k$-partite correlations,
\begin{eqnarray}\label{genuinek}
D^k(\rho_N):=D^{k-1\rightarrow N}(\rho_N)-D^{k\rightarrow N}(\rho_N).
\end{eqnarray}
As expected, the total correlations are given by the distance to the set of statistically independent $N$-partite states, which equals the sum of the correlations of any order:  $D^{1\rightarrow N}(\rho_N)=\min\limits_{\sigma \in P_1} D(\rho_N,\sigma)= \sum_k D^k(\rho_N)$.
 For example, the genuine bipartite correlations in a tripartite state  are computed as the difference between   total correlations and  genuine tripartite correlations,  $D^2(\rho_3)=D^{1\rightarrow 3}(\rho_3)- D^{2\rightarrow 3}(\rho_3)$, which are non-zero if and only if the state is not factorizable with respect to any bipartite cut.
 
  The minimization in Eq.~\ref{min} is  cumbersome for a generic distance $D$,  but significantly simplified by employing  the  relative entropy $S(\rho||\sigma)=-S(\rho)-\text{Tr}(\rho\log\sigma), \forall \rho,\sigma: \text{supp}~\rho\subseteq\text{supp}~\sigma,\  \infty\ \text{otherwise}, S(\rho)=-\text{Tr}(\rho\log\rho)$.   
In such a case, the closest product state to the global state is the product of its marginals, $\min\limits_{\otimes_{i=1}^m \sigma_{k_i}} S(\rho_N||\otimes_{i=1}^m \sigma_{k_i})=S(\rho_N||\otimes_{i=1}^m \rho_{k_i})=\sum_{i=1}^m S(\rho_{k_i})-S(\rho_N), \text{Tr}_{N-k}\rho_N=\rho_k$     \cite{modi,szalay}. 
Correlations of order higher than $k$ are then given by the distance to  $\tilde{P}_{k}\subset P_k,   \tilde{P}_{k}:=\left\{ \otimes_{j=1}^m \rho_{k_j}, \sum_{j=1}^m k_j=N, k=\max\{k_j\} \right\}$.  Therefore,  the genuine $k$-partite correlations  are measured by
\begin{eqnarray}\label{entk}
S^{k}(\rho_N)=S^{k-1\rightarrow N}(\rho_N)-S^{k\rightarrow N}(\rho_N).
\end{eqnarray}
 For systems invariant under subsystem permutations,  the subadditivity of the von Neumann entropy, $S(\rho_i)+S(\rho_j)\geq S(\rho_{ij}),$ makes the closest product state $\tilde{\rho}_N^k$ to be the most ``compact'' one, being the tensor product state  of $\lfloor N/k \rfloor$ clusters of  $k$ subsystems and a cluster of $N -\lfloor N/k \rfloor k= N \bmod k$ subsystems,  $S^{k\rightarrow N}(\rho_N)= \lfloor N/k\rfloor S(\rho_k)+ (1-\delta_{N \bmod k,0})S(\rho_{N \bmod k})-S(\rho_N)$.
 

We now verify the consistency of the framework. We identify  reasonable properties characterizing measures of $k$-partite correlations, applicable for any order $k$,  by generalizing the ones proposed for $N$-partite correlations \cite{bennett}.  We show that the quantities in   Eqs.~(\ref{min},\ref{genuinek}), for any distance $D$, satisfy the criteria 0D-4D  of invariance and monotonicity  under local operations and changes in the  system partition. We also prove that, by adopting the relative entropy, stronger yet desirable constraints 0S-5S are met:  \\

0D-0S) {\it The measures of $k$-partite correlations  are faithful}.  They are  non-negative, $D^k(\rho_N)\geq0$, and vanish if and only if the state does not have $k$-partite correlations. \\

1D) {\it Adding a disjoint $n$-partite system,  ${\cal S}'_{N+n}:={\cal S}_N\cup {\cal S}_n,$ cannot create correlations of order higher than $n$.} If the state of ${\cal S}_N$ does not have correlations of order higher than $n, \rho_N=\otimes_{j=1}^m \sigma_{k_j}, \sum_{j=1}^m k_j=N, n\geq\max\{k_j\}$, then  the state of   ${\cal S}'_{N+n}$ is $\rho_N\otimes\rho_n$, which does not have  correlations of order higher than $n$, the largest factor of the product being still a state of  $n$ or fewer subsystems. Thus, $D^{n\rightarrow N+n}(\rho_{N+n})= D^{n\rightarrow N}(\rho_N)=0$.  

1S) {\it  Adding a disjoint $n$-partite system cannot {\it increase}  correlations of order higher than $n$.} 
 One has $S^{n\rightarrow N}(\rho_N) =S(\rho_N||\tilde{\rho}_N^n)=S(\rho_N\otimes \rho_n||\tilde{\rho}_N^n\otimes\rho_n)\geq S(\rho_N\otimes \rho_n||\tilde{\rho}_{N+n}^n)=S^{n\rightarrow N+n}(\rho_{N+n})$. For example, given $N=3$, adding a bipartite system, $n=2,$ cannot increase the tripartite correlations.  \\
 
 2D-2S) {\it  Local  CPTP maps $\Pi_i\Phi_{[i]}, \Phi_{[i]}=I_1\otimes\ldots \Phi_i\otimes\ldots\otimes I_N,$   cannot create correlations of any order $k$,  and cannot increase the amount of correlations higher than any order $k$. }  Local operations do not change the tensor product structure of a state, $\rho_N \in P_{k} \Rightarrow \Pi_i\Phi_{[i]}(\rho_N)\in P_{k}, \rho_N \not\in P_{k} \Rightarrow \Pi_i\Phi_{[i]}(\rho_N)\not\in P_{k},$ so they cannot create  correlations of any order, $D^{k}(\rho_N)=0\Rightarrow D^{k}(\Pi_i\Phi_{[i]}(\rho_N))=0, \forall k$. Contractivity under CPTP maps guarantees  $D^{k\rightarrow N}(\rho_N)\geq D^{k\rightarrow N}(\otimes_i\Phi_{[i]}(\rho_N)), \forall k$.  This also implies   monotonicity of the highest order of non-zero correlations, $D^{\tilde{k}}(\rho_N)\geq D^{\tilde{k}}( \Pi_i\Phi_{[i]}(\rho_N)), \tilde{k}=\max\{k\}: D^{k\rightarrow N}(\rho_N) =0$.  Note that an operation  on a cluster of $n$ subsystems $\Phi_n, n>1,$ can create correlations of order up to $k+n-1$  from  already existing $k$-partite correlations. A state with non-zero $k$-partite correlations reads   $\otimes_{j=1}^m \sigma_{k_j}, \sum_{j=1}^m k_j=N, \max\{k_j\}\geq k$. A map $\Phi_n$ jointly applied to one subsystem of the largest cluster ${\cal S}_{\max\{k_j\}}$ and other  $n-1$ subsystems generates correlations of order up to $\max\{k_j\}+n-1$. For example,  the $k$-qubit  state $\ket{a_k}= a \ket{0}^{\otimes k}+\sqrt{1-a^2}\ket{1}^{\otimes k}, a \in(0,1),$ has genuine $k$-partite correlations $S^k(\ket{a_k})=2 (a^2\log{a^2}+(1-a^2)\log(1-a^2))$. Correlating the state with an ancillary target qubit by  a  CNOT gate creates the  state  $\ket{a_{k+1}}= a \ket{0}^{\otimes (k+1)}+\sqrt{1-a^2}\ket{1}^{\otimes (k+1)}$, which has $k+1$-partite correlations,  $S^k(\ket{a_{k+1}})= S^k(\ket{a_{k}})$. \\
 
 3D) {\it Partial trace of $n$ subsystems cannot increase correlations of order higher than $k<N-n$.}  Let   $\tilde{\rho}_N^k$ be the closest $N$-partite state with up to $k$-partite correlations. Then,  by  contractivity of the distance function, $D^{k\rightarrow N}(\rho_N)=D(\rho_N,\tilde{\rho}^k_N)\geq D(\rho_{N-n}, \text{Tr}_{n} \tilde{\rho}^k_N)\geq D^{k\rightarrow N-n}(\rho_{N-n})$.
 
3S) {\it Partial trace of $N-k$ subsystems can create up to $k$-partite correlations from existing $N$-partite correlations.} Let us consider the classical $N$-bit state $\rho^c_N=(\ket{0}\bra{0}^{\otimes N}+\ket{1}\bra{1}^{\otimes N})/2,$ which has $N$-partite correlations. The marginal state $\text{Tr}_{N-k}\rho^c_N=(\ket{0}\bra{0}^{\otimes k}+\ket{1}\bra{1}^{\otimes k})/2, \forall k,$ has genuine $k$-partite correlations. Contractivity ensures  $S^N(\rho_N)\geq S^k(\rho_k)$. The property is then  proven, as $k$-partite correlations are not necessarily present in $\rho_N^c$. One has  $S^k(\rho_N^c)=\lfloor N/(k-1) \rfloor- \lfloor N/k \rfloor+\delta_{N\bmod k,0 }-\delta_{N\bmod (k-1),0}=\lceil N/(k-1)\rceil- \lceil N/k\rceil$. 
For example, given $N=5$, the global state does not have genuine $4$-partite correlations, $S^2(\rho_5^c)=2, S^3(\rho_5^c)=S^5(\rho_5^c)=1, S^4(\rho_5^c)=0$. Indeed, the state $\rho^c_3\otimes \rho^c_2$ has more information about the global state $\rho^c_5$ than  $\rho^c_2\otimes \rho^c_2\otimes \rho^c_1$, thus there are genuine $3$-partite correlations, but the state $\rho^c_4\otimes \rho^c_1$ is  not more informative than the 3-vs-2 product (the relative entropy distance to the global state is equal). Genuine 4-partite correlations are distilled by tracing away one subsystem, $S^4(\rho_4^c)=1$.\\
%

4D-4S) {\it Distilling  $n$ subsystems by fine graining cannot create correlations of order higher than $k+n,$ for any $k$.}  Fine graining  a subsystem into a cluster of $n$ subsystems, ${\cal S}_{[i]}\rightarrow {\cal S}_{i'}=\{{\cal S}_{[i_j]}\}, j=1,\ldots,n+1$, changes the system partition  into ${\cal S}'_{N+n}=\{{\cal S}_{[1]}, {\cal S}_{[2]}, \ldots, {\cal S}_{[i-1]},  \{{\cal S}_{i_j}\},  {\cal S}_{[i+1]}, \ldots,  {\cal S}_{[N]}\}$.  If the state of the system ${\cal S}_N$ has  correlations  of order up to $k$, $\rho_N=\rho_{k\leq N}\otimes(\otimes_{l_j\leq k}\rho_{l_j}), \sum_j l_j =N-k$, the fine-graining map  creates at most  correlations of  order $k+n$, $\rho_{N+n}=\rho_{k+n\leq N+n}\otimes(\otimes_{l\leq k}\rho_l)$. Hence, $D^{k+n \rightarrow N+n}(\rho_{N+n})=D^{k\rightarrow N}(\rho_N)=0$.\\

5S) {\it Total correlations are superadditive.} It is given a coarse grained partition $\{{\cal S}_{k_1}, {\cal S}_{k_2},\ldots, {\cal S}_{k_l}\},  \sum_{j=1}^l k_j=N $. The total correlations in each cluster ${\cal S}_{k_j}=\{{\cal S}_{[\sum_{m=1}^{j-1}k_{m}+1]},{\cal S}_{[\sum_{m=1}^{j-1}k_{m}+2]},\ldots, {\cal S}_{[\sum_{m=1}^{j-1}k_{m}+k_j]}\}$
 are quantified by the  multi-information between the single subsystems forming the cluster, $S^{1\rightarrow k_j}(\rho_{k_j})=\sum_{n=1}^{k_j}  S(\rho_{[\sum_{m=1}^{j-1}k_{m}+n]})-S(\rho_{k_j})$, a non-negative extension of the bipartite mutual information \cite{wata,lindblad,horod}.  Exploiting  subadditivity, one has $\sum_{j=1}^l S^{1\rightarrow k_j}(\rho_{k_j})=\sum_{i=1}^N S(\rho_{[i]})-\sum_{j=1}^{l} S(\rho_{k_j})\leq \sum_{i=1}^N S(\rho_{[i]})-S(\rho_N) = S^{1\rightarrow N}(\rho_N)$, where the inequality is saturated for product states $\rho_{N}=\otimes_j \rho_{k_j}$. 
That is, the sum of the total correlations in each cluster  is  upper bounded by the total correlations in the global system.  For product states, subadditivity also implies additivity   for correlations higher than $k$, for every $k$,  $\rho_{N}=\otimes_j \rho_{k_j} \Rightarrow\tilde\rho_N^k=\otimes_{j} \tilde\rho_{k_j}^k \Rightarrow S^{k\rightarrow N}(\rho_N)=\sum_j S^{k\rightarrow k_j}(\rho_{k_j})$. \\
 
While being intuitive and simple to phrase, the discussed properties are not met by heavily employed measures and indicators of multipartite correlations. 
Covariances  of local observables $O_{[i]}$,  $\langle \Pi_i O_{[i]}\rangle_{\rho_N} -\Pi_i\langle O_i\rangle_{\rho_N}$,  do  not satisfy such criteria. They can vanish, for any choice of  $\{O_{[i]}\}$,  in presence of  classical and quantum multipartite correlations \cite{bennett,kaz,wal}.  An alternative correlation  witness measures the ability of multipartite systems to extract work from local environments \cite{bennett}, yet being still unproven whether the quantity satisfies properties 0D-4D.  Another measure of correlations above order $k$ is the (relative entropy) distance of the global state to the  state with maximal  von Neumann entropy, among the ones with the same marginal states of $k$ subsystems, $S(\rho_N||\bar{\sigma}_N^k), \bar{\sigma}_N^k:=\max\limits_{\sigma_N: \text{Tr}_{N-k}\sigma_N=\rho_k} S(\sigma_N)$ \cite{amari,linden,bialek,zhou,jost}.  Remarkably,   independent lines of thinking converged to the very same definition.  However, such measure, as well as a related one given by the trace norm of the cumulant of the state \cite{zhou3,wal}, violates contractivity under local operations in both classical and quantum scenarios \cite{comp2,zhou2}. This happens because local operations do change a state whilst preserving its tensor product structure, thus changing the set of states with  the same $k$-marginals. \\



\begin{table*}[t]
{\footnotesize
\begin{tabular}{|c|c|c|c|c|} 
\hline
\textbf{$\mathbf{\rho_N, N}$ even}   & $\mathbf{S^{k}, k<N}$& $\mathbf{S^{N}}$&$\mathbf{S^{1\rightarrow N}}$   & $\mathbf{W_S, \omega_k=k-1}$ \\
\hline
$[(\ket{00}\bra{00}+\ket{11}\bra{11})/2]^{\otimes N/2}$  &$N/2 \delta_{k,2}$&0&$N/2$& $ N/2$\\
\hline
$(\ket{0}\bra{0}^{\otimes N}+\ket{1}\bra{1}^{\otimes N})/2$  &
$\lceil N/(k-1) \rceil -\lceil N/k\rceil $ 
 &1&$N-1$& $  \sim 1.13 N\log N-N$\\
\hline
$ [\frac{1}{\sqrt{2}}(|00\rangle +|11\rangle)]^{\otimes N/2}$  &$N\ \delta_{k,2}$ & 0     & $N$  &     $N$    \\
\hline
$\frac{1}{\sqrt{2}}(|0\rangle^{\otimes N}+|1\rangle^{\otimes N})$  &$\lceil N/(k-1) \rceil -\lceil N/k\rceil $ 
 &  2  & $N$      & $ \sim 1.13 N \log N $     \\
\hline
 $\frac{1}{\sqrt{\binom{N}{1}}}\sum_i {\cal P}_i ( |0\rangle^{\otimes (N-1)}\otimes|1\rangle)$   &$f^k_{\text{D},1}$ &  \ $2/N(h(N)-h(N-1))\sim 0$   &  $h(N)-h(N-1)\sim\log N$   &   $\sim 2.61 N$     \\      
\hline 
$\frac{1}{\sqrt{\binom{N}{N/2}}}\sum_i {\cal P}_i ( |0\rangle^{\otimes N/2}\otimes|1\rangle^{\otimes N/2})$   &$f^k_{\text{D},N/2} $&   $2$   &  
  $ N$       &   $\sim  0.01 N^2 $ \\            
\hline      
$\sum_{i=1}^d \ket{i}\bra{i}^{\otimes N}/d$  & $(\lceil N/(k-1) \rceil -\lceil N/k\rceil) \log d$    &  $\log d$    & $(N-1)\log d$      & $\sim (1.13 N\log N-N)\log d$ \\ 
\hline
$(\sum_{i=1}^d \ket{ii}/\sqrt d)^{\otimes N/2}$  &  $N \log d\ \delta_{k,2}$ & 0    & $N \log d$      &   $N\log d$ \\       
\hline  
\end{tabular}
}
\caption{Genuine $k$-partite correlations, $N$-partite correlations, total correlations, and weaving (asymptotic scaling for $N\rightarrow \infty$) for: product of  $N/2$ maximally correlated two-bit states; maximally correlated  $N$-bit state; product of $N/2$ Bell states; $N$-partite GHZ state (the expressions hold for $N$ odd as well);  $N$-partite Dicke states  with one  and $N/2$ excitations,  being $h(x)=x \log x,$ and the functions $f_{\text{D},1}, f_{\text{D},N/2}$ given in Ref.~\cite{epaps};  maximally correlated $N$-partite classical state of dimension $d$;   product of $N/2$ maximally entangled two-qudit states.}
\label{table}
\end{table*}
 
{\it Ranking correlation patterns by weaving --} Having determined how to quantify genuine multipartite correlations, one observes that equally correlated states, in terms of total correlations, can display different values of correlations for some order $k$, and thus different properties.   Assuming $N$ even, a  product of $N/2$  Bell states, e.g. $|\psi_{N/2}\rangle=[1/\sqrt{2}(\ket{00}+\ket{11})]^{\otimes N/2}$, has the same total correlations of the $N$-partite GHZ state $\ket{\text{GHZ}_N}=\frac{1}{\sqrt{2}}(|0\rangle^{\otimes N}+|1\rangle^{\otimes N}),$ as measured by the relative entropy,  $S^{1\rightarrow N}(|\psi_{N/2}\rangle)= S^{1\rightarrow N}(\ket{\text{GHZ}_N})=N,$ whilst the latter exhibits correlations of higher order. 
 On the same hand,  the highest order of  correlations is not sufficient  to describe multipartite states. Both the GHZ and the  $N/2$-excitation Dicke state $\frac{1}{\sqrt{\binom{N}{N/2}}}\sum_i {\cal P}_i ( |0\rangle^{ \otimes N/2}\otimes |1\rangle^{\otimes N/2})$, where the sum is over the permutations $\{{\cal P}_i\}$ of the   group $\text{Sym}(N)$, have two bits of $N$-partite correlations. Yet, they have different uses for information processing \cite{dicke,dicke2}, and it is impossible to transform them into each other by local operations and classical communication \cite{cirac}.\\
We introduce the concept of weaving to rank classical and quantum multipartite states by a single index, overcoming such ambiguities. The idea is to construct a consistent information-theoretic descriptor of correlation patterns by counting well-defined  genuine multipartite correlations of every order. A weaving measure is built as the weighted sum of  multipartite correlations,  
\begin{eqnarray}
	W_D(\rho_{N})= \sum_{k=2}^{N}\omega_k D^k(\rho_N) = \sum_{k=1}^{N-1}\Omega_k D^{k\rightarrow N}(\rho_N),
\end{eqnarray}
				where $\omega_k=\sum\limits_{i=1}^{k-1}\Omega_i, \omega_k \in \mathbb R^+$.  For any function $D$, a weaving measure is contractive under local operations and partial trace,  $W_D(\rho_N)\geq W_D(\Pi_i \Phi_{[i]}(\rho_N)), W_D(\rho_N)\geq W_D(\rho_k)$, as it is a sum of contractive quantities (properties 2D, 3D).  The relative entropy of weaving $W_S(\rho_N)$ is also additive, $W_S(\otimes_j\rho_{k_j})=\sum_j W_S(\rho_{k_j})$, being a sum of additive terms (property 5S). Weaving is then easy to compute too, being obtained by  global and marginal entropies. 

The choice of the weights determines the meaning  of a weaving measure.  For  $\omega_k=1, \forall k$,  it is a measure of total correlations.  For $\omega_{l}=\delta_{kl}, \forall l,$ it  quantifies genuine $k$-partite correlations. As observed, computing correlations is not sufficient to discriminate different multipartite states.   Thus, we study correlation scaling. That is,  how the information about the global system scales  by accessing partitions containing clusters of increasing size. This is captured, for example,  by choosing weights proportional to the correlation order.  We calculate the relative entropy measures of genuine multipartite correlations and  weaving, selecting $\omega_k=k-1\Rightarrow \Omega_i=1, \forall i$, for  highly correlated classical and quantum states of $N$ particles  (Table~\ref{table}).  The quantity  unambiguously ranks  states with equal total correlations, or $N$-partite correlations. As expected, the weaving of  states in tensor product form, e.g. the Bell state products, scales linearly $O(N)$ with the number of particles. Indeed,  the correlations in the global state are the sum of the correlations in each product factor. The GHZ state shows super-linear  scaling $O(N \log N)$ instead.  However, the highest asymptotic value $O(N^2)$ for $N$ qubits is found in the $N/2$ excitation Dicke state. Such state has non-zero correlations at any order, $f^k_{\text{D},N/2}\neq 0, \forall k$ \cite{epaps}, while the GHZ state has zero correlations whenever $\lceil N/(k-1) \rceil =\lceil N/k\rceil $.  Weaving is proportional to the logarithm of  the subsystem dimension $d$.  \\ 
The concept of weaving solves issues emerged in previous studies.   A measure of   ``neural complexity'' was proposed  to study correlation scaling between binary variables \cite{tse}. The quantity, which we generalize to  the quantum scenario, reads $C(\rho_N)=\sum_{k=1}^{N-1} k/N\  S^{1\rightarrow N}(\rho_N)-\langle S^{1\rightarrow k}(\rho_k)\rangle$, where the average term is calculated over the $\binom{N}{k}$  clusters of $k$ subsystems ${\cal S}_k$.   A geometric lower bound is given by  the weighted distances to the  entropy maximisers with same $k$-marginals,  $C(\rho_N)\geq C^g(\rho_N)=\sum_k k/N S(\rho_N,\bar{\sigma}_N^k)$ \cite{comp1,comp2,jost,jost2,attempt,revcomp}.  
 The interest in complexity measures was spurred by the  association with enhanced neuronal activity, evaluating  the functionality of  equally correlated  neural networks. This generated a  debate about whether complexity is the resource governing information transmission in the brain \cite{tonrev}. Such quantities have been also applied to study chaotic systems and cellular automata \cite{comp2}.  
 Yet,  complexity measures fall short as benchmarks of multipartite correlations.  The neural complexity is not additive under tensor products,  e.g. $C(\rho_{2}\otimes \rho_1)=4/3  C(\rho_2)$, while the geometric complexity is not contractive under local operations, 
   nor under partial trace, requiring non-analytical methods to be computed \cite{comp2,attempt,note0}.     \\


   {\it Conclusion  --}
   We proposed a consistent information-theoretic definition of genuine multipartite correlations, and described how to quantify them. While we did not discuss the distinction between  classical and quantum correlations,  our result suggests a strategy to  characterize genuine multipartite  {\it quantum}   correlations, an open question despite recent progresses \cite{multi3,giorgi,modi,paula}. Having defined  $k$-partite correlations as in Eq.~\ref{min},  classical and quantum contributions  can be  identified via the method employed for total correlations \cite{modi}, then studying quantum correlations on their own. \\
 We also introduced weaving, a descriptor of correlation patterns.    Weaving is an  alternative to complexity measures, i.e. a measure of how hard is to determine the properties of a system from knowing its parts \cite{aps}, which satisfies desirable constraints.  An important question to address is its operational meaning. Specifically, the quantum contribution to the weaving of a state may  be a further computational resource. This would confirm the intuition that interplaying complexity science and (quantum) information theory can advance both disciplines \cite{biamonte}.   


\section*{Acknowledgments}
We thank  O. Guehne,  J. H. Reina, V. Vedral, C. Viviescas, and B. Yadin for fruitful discussions. D. G. is supported by the  UK Engineering and Research Council (Grant No. EP/L01405X/1) and the Wolfson College, University of Oxford.  T. T.  is supported by the Foundational Questions Institute (fqxi.org), Physics of the Observer Programme (Grant No. FQXi-RFP-1601).  C. E. S. Thanks Colciencias for a fellowship, CIBIOFI and Universidad de C\'{o}rdoba.

   \clearpage
\onecolumngrid
\renewcommand{\bibnumfmt}[1]{[A#1]}
\renewcommand{\citenumfont}[1]{{A#1}}
\setcounter{page}{1}
\setcounter{equation}{0}

\appendix*

\section{Appendix}

\section*{Genuine $k$-partite correlations of Dicke states}
The Dicke states with one and $N/2$ excitations read  
\begin{eqnarray}
\ket{\text{D},1}&=&\frac{1}{\sqrt{\binom{N}{1}}}\sum_i {\cal P}_i \left( |0\rangle^{\otimes (N-1)}\otimes|1\rangle\right), \\
\ket{\text{D},N/2}&=&\frac{1}{\sqrt{\binom{N}{N/2}}}\sum_i {\cal P}_i \left( |0\rangle^{\otimes N/2}\otimes|1\rangle^{\otimes N/2}\right).
\end{eqnarray}
Since they are invariant under subsystem permutations, we can employ the compact expression given in the main text:
\begin{eqnarray}
S^{k\rightarrow N}(\rho_N)= \lfloor N/k\rfloor S(\rho_k)+ (1-\delta_{N \bmod k,0})S(\rho_{N \bmod k})-S(\rho_N).
\end{eqnarray}
Then, the amount of genuine $k$-partite correlations of the Dicke states  is given by
\begin{eqnarray}
f^k_{\text{D},1}:&=&S^{k-1\rightarrow N}(\ket{\text{D,1}})-S^{k\rightarrow N}(\ket{\text{D},1}),\nonumber \\
S^{k\rightarrow N}(\ket{\text{D},1})&=& \frac{\left\lfloor \frac{N}{k}\right\rfloor  \left((k-N) \log \left(1-\frac{k}{N}\right)-k \left(\log \left(\frac{k \left\lfloor \frac{N}{k}\right\rfloor }{N}\right)-\log
   \left(1-\frac{k \left\lfloor \frac{N}{k}\right\rfloor }{N}\right)+\log \left(\frac{k}{N}\right)\right)\right)-N \log \left(1-\frac{k \left\lfloor \frac{N}{k}\right\rfloor
   }{N}\right)}{N}, \\
f^k_{\text{D},N/2}:&=&S^{k-1\rightarrow N}(\ket{\text{D},N/2})-S^{k\rightarrow N}(\ket{\text{D},N/2}),\nonumber \\
S^{k\rightarrow N}(\ket{\text{D},N/2})&=&\left\{\left\lfloor \frac{N}{k}\right\rfloor  \sum _{i=0}^k -\frac{\binom{k}{i} \binom{N-k}{\frac{N}{2}-i} \log \left(\frac{\binom{k}{i}
   \binom{N-k}{\frac{N}{2}-i}}{\binom{N}{\frac{N}{2}}}\right)}{\binom{N}{\frac{N}{2}}}+\sum _{i=0}^{N-k \left\lfloor \frac{N}{k}\right\rfloor } -\frac{\binom{k \left\lfloor
   \frac{N}{k}\right\rfloor }{\frac{N}{2}-i} \binom{N-k \left\lfloor \frac{N}{k}\right\rfloor }{i} \log \left(\frac{\binom{k \left\lfloor \frac{N}{k}\right\rfloor }{\frac{N}{2}-i}
   \binom{N-k \left\lfloor \frac{N}{k}\right\rfloor }{i}}{\binom{N}{\frac{N}{2}}}\right)}{\binom{N}{\frac{N}{2}}}\right\}.
\end{eqnarray}
One can verify that the Dicke state with $N/2$ excitations displays correlations at any order, for arbitrary $N$, $f^k_{\text{D},N/2}\neq 0, \forall N, k$.


\begin{thebibliography}{99}

\bibitem{cover}T. Cover and J. Thomas, {\it Elements of Information Theory}, Wiley (1991).

\bibitem{badii}R. Badii and A. Politi, {\it Complexity: Hierarchical Structure and Scaling in Physics.} Cambridge University Press (1997).
\bibitem{condmat}J. P. Sethna, {\it Statistical Mechanics: Entropy, Order Parameters, and Complexity}, Oxford University Press (2006).

\bibitem{wilde}M. Wilde, {\it Quantum Information Theory}, Cambridge University Press (2013).

\bibitem{many}L. Amico, R. Fazio, A. Osterloh, V. Vedral, Rev. Mod. Phys. {\bf 80}, 517 (2008).


\bibitem{ent}R. Horodecki, P. Horodecki, M. Horodecki, and K. Horodecki,  Rev. Mod. Phys. {\bf 81}, 865 (2009).

\bibitem{bennett}C. H. Bennett, A. Grudka, M. Horodecki, P. Horodecki, and R. Horodecki, Phys. Rev. A  {\bf 83}, 012312 (2011).

\bibitem{multi1}O. Guehne, G. T\'oth,  Phys. Rep. {\bf 474}, 1 (2009).
\bibitem{multi2}M. Huber, F. Mintert, A. Gabriel, and  B. C. Hiesmayr,  Phys. Rev. Lett. {\bf 104}, 210501 (2010).
\bibitem{multi3} F. Levi and F. Mintert, Phys. Rev. Lett. {\bf 110}, 150402 (2013).
\bibitem{multi4}L. Pezz\'e and A. Smerzi,  Phys. Rev. Lett. {\bf 110}, 163604 (2013).

\bibitem{multi6}K. Schwaiger, D. Sauerwein, M. Cuquet, J. I. de Vicente, and B. Kraus, Phys. Rev. Lett. {\bf 115}, 150502 (2015).
\bibitem{szalay}S. Szalay, Phys. Rev. A {\bf 92}, 042329 (2015).

\bibitem{multi5} D. Girolami and B. Yadin,  Entropy {\bf 19}, 124 (2017).


 \bibitem{multi7}I. Apellaniz, M. Kleinmann, O. Guehne, and G. T\'oth,	Phys. Rev. A {\bf 95}, 032330 (2017).



\bibitem{comp1}N. Ay, E. Olbrich, N. Bertschinger, and J.  Jost, Chaos  {\bf 21}, 037103 (2011).
\bibitem{comp2}T. Galla and O. Guehne,  Phys. Rev. E  {\bf 85}, 046209 (2012).
\bibitem{cirac}W. D\"{u}r, G. Vidal, and J. I. Cirac, Phys. Rev. A {\bf 62}, 062314 (2000). 
\bibitem{acin}A. Ac\`{i}n, D. Bruss, M. Lewenstein, and A. Sanpera, Phys. Rev. Lett. {\bf 87}, 040401 (2001).
\bibitem{nine}F. Verstraete, J. Dehaene, B. De Moor, and H. Verschelde, Phys. Rev. A {\bf 65}, 052112 (2002).


\bibitem{lloyd1}P. Zanardi, D. Lidar and S. Lloyd,  Phys. Rev. Lett. {\bf 92}, 060402
(2004).
\bibitem{viola}H. Barnum, E. Knill, G. Ortiz, R. Somma, and L. Viola, Phys. Rev. Lett. {\bf 92}, 107902 (2004).

\bibitem{briegel}O. Guehne, G. T\'{o}th, and H. J. Briegel, 	New J. Phys. {\bf 7}, 229 (2005).


\bibitem{modi}K. Modi, T. Paterek, W. Son, V. Vedral, and M. Williamson, Phys. Rev. Lett. {\bf 104}, 080501 (2010).


\bibitem{wata}S. Watanabe, IBM J. of Res. and Dev. {\bf 4}, 66 (1960). 

\bibitem{lindblad} G. Lindblad, Commun.
Math. Phys. {\bf 33}, 305 (1973).

\bibitem{horod}R. Horodecki, Phys. Lett. A {\bf 187}, 145 (1994).

\bibitem{kaz}D. Kaszlikowski, A. Sen, U. Sen, V. Vedral, and
A. Winter, Phys. Rev. Lett. {\bf 101}, 070502 (2008); Z. Walczak, Phys. Rev. Lett. {\bf 104}, 068901 (2010); D. Kaszlikowski, A. Sen, U. Sen, V. Vedral, and A. Winter, Phys. Rev. Lett. {\bf 104}, 068902 (2010);
 C. Schwemmer, L. Knips, M. C. Tran, A. de Rosier, W. Laskowski, T. Paterek, and H. Weinfurter, Phys. Rev. Lett. {\bf 114}, 180501 (2015).

\bibitem{wal}Z. Walczak, Phys. Lett. A {\bf 374}, 3999 (2010).


\bibitem{amari}S. Amari,  IEEE Trans. IT  {\bf 47}, 1701  (2001).

\bibitem{linden}N. Linden and W. K. Wootters, Phys. Rev. Lett. {\bf 89}, 277906  (2002).

\bibitem{bialek}E. Schneidman, S. Still, M. J. Berry II, and W. Bialek, Phys. Rev. Lett. {\bf 91}, 238701 (2003).

\bibitem{zhou}D. L. Zhou, Phys. Rev. Lett. {\bf 101}, 180505 (2008).

\bibitem{jost}E. Olbrich, T, Kahle, N. Bertschinger, N. Ay, and J. Jost, Eur. Phys. J. B {\bf 77}, 239 (2010).
\bibitem{zhou3}D. L. Zhou, B. Zeng, Z. Xu, and L. You,  Phys. Rev. A {\bf 74}, 052110 (2006).

\bibitem{zhou2}D. L. Zhou,  Phys. Rev. A {\bf 80}, 022113 (2009).





\bibitem{dicke}J.K. Stockton, J.M. Geremia, A.C. Doherty, and H. Mabuchi, Phys. Rev. A {\bf 67}, 022112 (2003).

\bibitem{dicke2}G. T\'{o}th, J. Opt. Soc. Am. B {\bf 24}, 275 (2007).


\bibitem{epaps}See Supplementaries.



 \bibitem{tse}G. Tononi, O. Sporns, and G. M. Edelman,   Proc. Natl. Acad. Sci. USA {\bf 91}, 5033 (1994).

\bibitem{revcomp}N. Ay , E. Olbrich, N. Bertschinger, J. Jost, {\it Proceedings of ECCS06 (European Complex Systems Society, 2006)}. 

\bibitem{jost2}T. Kahle, E. Olbrich, J. Jost, and N. Ay,  Phys. Rev. E {\bf 79}, 026201  (2009).

\bibitem{attempt}S. Niekamp, T. Galla, M. Kleinmann, and O. Guehne,  J. Phys. A: Math. Theor.  {\bf 46}, 125301 (2013).

\bibitem{tonrev}G. Tononi, G. M.  Edelman, and O. Sporns,  Trends Cogn. Sci. {\bf 2}, 474 (1998).  
 
 \bibitem{note0}In Ref.~\cite{comp2}, the authors propose a complexity measure for classical systems which is contractive under local operations, but harder to compute. 









 














\bibitem{giorgi}G. L. Giorgi, B. Bellomo, F. Galve, and R. Zambrini, Phys. Rev. Lett. {\bf 107}, 190501 (2011).



\bibitem{paula}F. M. Paula, A. Saguia, Thiago R. de Oliveira, M. S. Sarandy, 	Europhys. Lett. {\bf 108}, 10003 (2014).



\bibitem{aps}To date (30 July 2017), the term ``complexity'' appeared 24472 times in  APS journals, with different means and meanings. To avoid further ambiguities, we adopt the term {\it weaving} to describe correlation patterns.


\bibitem{biamonte}M. De Domenico and J. Biamonte, Phys. Rev. X {\bf 6}, 041062 (2016).


 \end{thebibliography}
    \end{document}